# Precise surface temperature measurements at kHz-rates using phosphor thermometry to study flame-wall interactions in narrow passages

Anthony O. Ojo[1,*], David Escofet-Martin[1], Christopher Abram[2], Benoit Fond[3], Brian Peterson[1]

[1]Institute for Multiscale Thermofluids, The University of Edinburgh, Edinburgh, EH9 3FD, UK.

[2]Department of Mechanical and Aerospace Engineering, Princeton University, Princeton, NJ 08540 USA

[3]Institute of Fluid Mechanics and Thermodynamics, Otto-von-Guericke Universität, 39106 Magdeburg, Germany

*Corresponding author: anthony.ojo@ed.ac.uk

**Abstract**

The thermographic phosphor $ScVO_4:Bi^{3+}$ is used to obtain time-resolved surface temperature measurements with sub-°C precision at 5 kHz. Measurements are used to study transient heat loss and flame-wall interactions (FWI) within a dedicated narrow two-wall passage (crevice) in an optically accessible fixed volume chamber. This passage emulates a crevice relevant in many technical environments, where FWI is less understood due to lack of detailed measurements. Chemiluminescence (CH*) imaging is performed simultaneously with phosphor thermometry to resolve how the flame's spatiotemporal features influence the local surface temperature. $ScVO_4:Bi^{3+}$ is benchmarked against $Gd_3Ga_5O_{12}:Cr,Ce$, a common phosphor used at low-kHz rates in FWI environments. $ScVO_4:Bi^{3+}$ is shown to offer higher luminescence signal levels and temperature sensitivity as well as negligible cross dependence on the excitation laser fluence, improving the precision and repeatability of the wall temperature measurement. $ScVO_4:Bi^{3+}$ is further used to resolve transient heat loss for variations in crevice spacing and uniquely capture temperature transients associated with flame dynamics. Taking advantage of these precise surface temperature measurements the wall heat flux is calculated with crevice spacing of 1.2 mm, where flame extinction is prevalent. Wall heat flux and estimated quenching distance are reported for flames that actively burn or extinguish at the measurement location.

## 1.0. Introduction

Flame-wall interactions (FWI) and transient heat loss at gas/wall interfaces are important topics in the design of thermal machines, such as downsized internal combustion (IC) engines and gas turbines [1,2]. Downsized IC engines play a critical role for reduced fuel consumption and lower $CO_2$ [3]. Such engines are important in future scenarios as they are seen as a shared powertrain in hybrid vehicles and a secondary powertrain for electric vehicles. Engine downsizing, however, increases surface-to-volume ratios such that gases and flames are more exposed to walls, thereby increasing gas-wall interactions. This leads to increased gaseous heat loss to wall surfaces and flame quenching, which limit efficiency gains and increase unburned hydrocarbon (UHC) emissions [2,4,5].

FWI has been studied in detail using sophisticated laser diagnostics (e.g., [6–10]) and numerical simulations (e.g., [11–15]). The more detailed studies have been conducted for single-wall configurations, which provide ample space for experimental measurements. However, heat loss and flame quenching are most severe in two-wall passages as the gas/flame is more exposed to walls due to higher surface-to-volume ratios. Such passages exist as crevices or slots within technically relevant systems (e.g. piston crevice, annular slots, flame arrestor). Piston crevices, for example, are a significant contributor to thermal stratification and unburned hydrocarbons in IC engines [2,5,16–18]. Understanding heat losses and FWI in such regions is important to design cleaner, more efficient engine technology.

Measurements of FWI in two-wall passages are less common, primarily due to limited optical access in these passages. Within production engines, ion probes have been used to identify the presence of a flame within crevices [2,19–22]. While such probes have recorded quenching distances as narrow as 0.1 mm [19], they are unable to describe the FWI in detail. More recently, large-eddy simulations coupled with high-speed imaging in an optically accessible engine have revealed that a flame is not merely pushed into a piston crevice, but can actively burn within a piston crevice [23]. While piston crevice dimensions are slightly larger for optical engines than production engines, such studies emphasize the need for detailed wall and flame modelling within narrow passages. Experimental measurements of wall temperature and FWI quantities are needed to support such modelling efforts.

Phosphor thermometry provides effective means to measure surface temperature in FWI applications [24]. This technique exploits the temperature-dependent luminescence properties of ceramic materials doped with rare-earth or transition metals. These thermographic phosphors (TGP) are coated unto the surface of interest and excited by incident laser light. The luminescence of the TGP is detected, and its temperature is derived using either the spectral intensity ratio or the lifetime approach [25]. TGPs have been widely applied on single-wall surfaces within IC engines and gas turbines operating under fired conditions [8,26–31]. With the exception of a few works that report TGP findings at 1 kHz repetition rates [8,30,31], all TGP findings in FWI applications have been limited to 10 Hz recording rates. To resolve the highly transient flame-wall dynamics at short timescales (<100 μs) in two-wall passages, single-shot high-speed measurements at several kHz are required. For such kHz measurements, the lifetime of the TGP selected for thermometry needs to be fast and be adequately resolved by a photodetector.

Bismuth-doped vanadate phosphors, such as $ScVO_4:Bi^{3+}$, are promising phosphors for kHz repetition rates as their luminescence lifetime is less than 10 μs at room temperature [32,33]. Although the luminescence of $ScVO_4:Bi^{3+}$ was studied in the 1970's [34], it was only recently evaluated by Abram et al. for thermometry [33]. Its lifetime sensitivity was shown to be among the highest reported for lifetime-based thermometers over the 20-60°C temperature range with 2.2%/ °C. This temperature range is applicable for FWI under cold-wall conditions with moderate heat release [8,31,35,36]. Abram et al. [33] applied $ScVO_4:Bi^{3+}$ for fluid temperature imaging in a liquid flow at 10 Hz, achieving a single shot single pixel precision of 0.4°C. To the authors' knowledge, $ScVO_4:Bi^{3+}$ has not been reported for surface thermometry, and has not been reported at kHz repetition rates.

This work presents a unique application of $ScVO_4:Bi^{3+}$ to measure surface temperature and FWI quantities (wall heat flux and flame quenching distance) within a two-wall passage. $ScVO_4:Bi^{3+}$ measurements are applied at 5 kHz and combined with chemiluminescence (CH*) imaging to capture the transient flame-wall dynamics, including flame quenching. $ScVO_4:Bi^{3+}$ is first benchmarked against the phosphor $Gd_3Ga_5O_{12}:Cr,Ce$, which is a prominent phosphor used at low-kHz repetition rates in FWI environments [8,30,31]. The performance of both TGPs is evaluated in a two-wall crevice of an optically accessible combustion chamber, whose geometry and operation emulates that of a piston engine. The advantages of $ScVO_4:Bi^{3+}$ in this particular FWI environment are reported. $ScVO_4:Bi^{3+}$ is further used to investigate wall temperature transients relative to variations in crevice spacing and flame dynamics. Wall heat flux is calculated to estimate the quenching distance for flames that extinguish or propagate beyond the wall measurement location.

## 2.0. Experimental setup

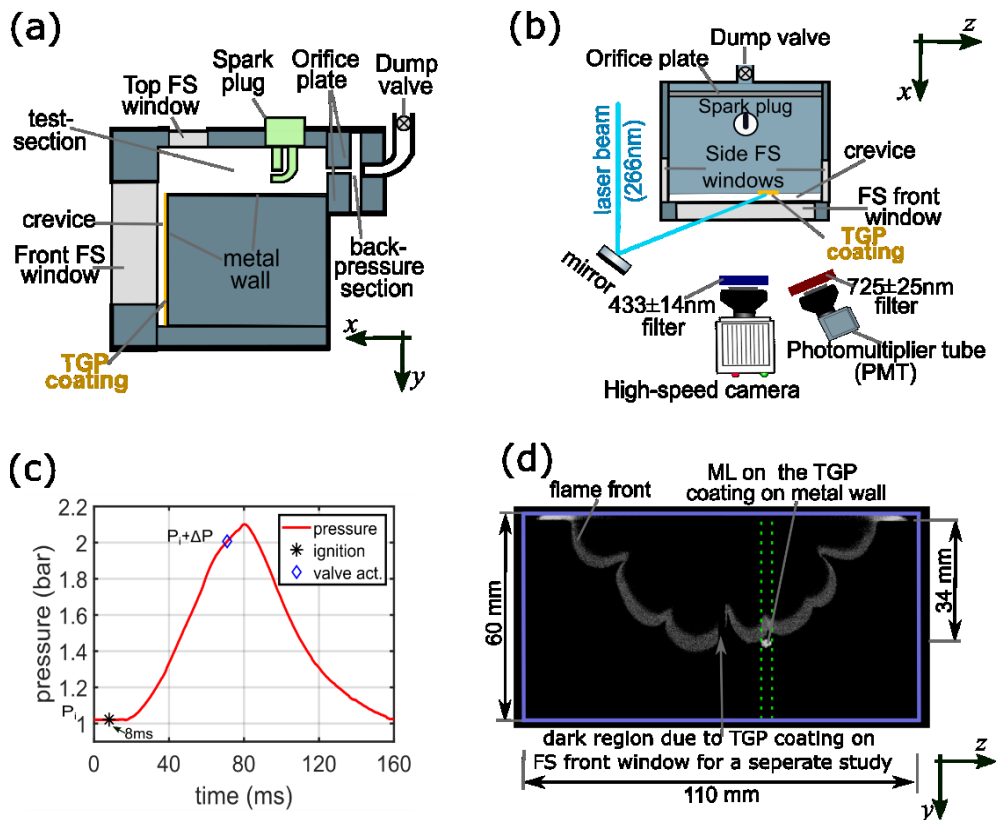

**Figure 1:** (a) Schematic of the FVC, (b) schematic of the experimental setup, (c) chamber's pressure-time curve for $P_i$ = 1 bar, (d) flame front image showing flame propagating downward into the crevice and the measurement location (ML). In (d), the blue outline is the boundary of the front window and the green-dash outline is the section used to track flame position.

### 2.1. Fixed volume combustion chamber (FVC)

Experiments are performed within an optically accessible fixed volume chamber (FVC). Figures 1a and b show the schematic of the FVC. The FVC features a test section (150 cm³) and a back-pressure section (6 cm$^3$). Separating these sections, there is a 6 mm thick orifice plate with 81 equidistant holes of 0.5 mm diameter. The test section emulates a simplified piston engine geometry at the top-dead center, including a crevice region. Optical access is provided by fused silica (FS) windows mounted at the front, top and two sides of the FVC. Metal components of the FVC are manufactured from 304 stainless steel.

The chamber operation is modelled after [37] and is demonstrated in [8,31]. The chamber is evacuated to 20 mbar using a vacuum pump. A homogeneous methane/air mixture is introduced into the FVC until an initial pressure ($P_i$) is reached, e.g., 1 bar (Fig 1c). The mixture is ignited via the spark plug. Heat release initiates an exponential pressure rise as the flame propagates towards the opposite end of the chamber before entering the crevice region. At a preselected pressure ($P_i + \Delta P$) e.g., 2 bar (Fig 1b), a dump-valve (8 ms response time) is actuated to evacuate the chamber. The exiting exhaust flow is choked via the orifice plate, providing an exponential pressure decay.

This work focuses on wall temperature measurements and flame front imaging within the two-walled crevice region of the FVC. Detailed measurements within the test-section (i.e. single-wall region) can be found in [8,31]. The crevice is 70 mm deep ($\Delta y$), and 158 mm wide ($\Delta z$). The crevice depth is similar in dimension to piston crevices in optically accessible piston engines [23,38]. The crevice region is characterised by two walls; one being the front FS window surface and the other being a metal wall (304 stainless steel). The distance between both walls ($\Delta x$) is the crevice spacing (CS). The front FS window placement can be adjusted for variable crevice spacings (0.5-5 mm). The front window ($\Delta y \times \Delta z$; 60 ×110 mm) provides optical access for the majority of the crevice region, as well as 2 mm ($\Delta y$) above the crevice within the test-section.

In this work, a variety of operating conditions involving variations in CS and $P_i$ was performed. Variations in these parameters greatly influence the heat transfer and FWI attributes in the crevice. The crevice spacings used in work are CS = 1.2, 2.0, and 3.5 mm. These CS prescribe the surface area to volume ratio in the crevice, which largely governs gas/wall heat exchange. The corresponding surface area to volume ratios ($SA/V$) are 1.69 mm$^{-1}$, 1.02 mm$^{-1}$ and 0.59 mm$^{-1}$, respectively. Initial pressure is used to dictate the depth the flame penetrates into the crevice. The flame penetrates further into the crevice for higher $P_i$ and larger CS. The pressures used in this work are $P_i$ = 1 bar, 1.4 bar and 2 bar. The preselected pressure ($P_i + \Delta P$) for dump valve actuation was 2.1 bar, 2.85 bar and 4.1 bar, respectively. The chamber walls are not externally cooled or heated in this work. Unless otherwise stated, methane/air mixture with equivalence ratio, $\phi$ = 1.0 was used for experiments, and the mixture was ignited at 8 ms after the start of the experiments (e.g. see Fig. 1c).

It is recognized that the CS in this work is larger than those in operating engines. The selected CS are chosen for fundamental FWI studies in two-wall passages. Such passage dimensions are more relevant to micro-combustors or crevices within larger marine engines.

## 2.2. Phosphor thermometry

In this work, two thermographic phosphors are evaluated for surface thermometry: (1) chromium-doped (2 mol%), cerium co-doped (0.034 mol%), gadolinium gallium garnet (Gd$_3$Ga$_5$O$_{12}$:Cr,Ce; hereafter called GGG) and (2) bismuth-doped (1 mol%) scandium vanadate ($ScVO_4:Bi^{3+}$; hereafter called ScVO). The GGG phosphor was obtained commercially (Phosphor Technology Ltd, UK). Two separate ScVO phosphor samples were employed in this work. The first is a sample synthesised in Ref [33]. This sample was used for the measurements reported in Sect. 3.0 and Sect. 4.1. The other ScVO sample was sourced commercially (Phosphor Technology Ltd, UK) and was used for measurements reported in Sect. 4.2. Both ScVO samples exhibit comparable luminescence characteristics including brightness, temperature sensitivity, and excitation fluence dependence.

Each respective phosphor (GGG or ScVO) was mixed with a temperature resistant HPC binder (ZYP coatings) and applied as a ~ 60 mm × 4 mm ($\Delta y \times \Delta z$) thin coating, which spanned the height of the crevice region as shown in Fig. 1a and b. The coating thickness was < 10 μm ($\Delta x$). Experiments were performed for GGG, after which the coating was removed, the surface cleaned and the coating strip with ScVO was applied in the same location as GGG. The coating was excited by an Edgewave Nd:YAG laser (266 nm). For GGG, the laser operated at 1 kHz, while for ScVO it was possible to run the laser at 5 kHz due to ScVO's shorter lifetime. Using a pinhole, the laser beam covered an area of about 10 mm$^2$ on the coating. The measurement location (ML) can be specified at a height ($\Delta y$) from the top of the crevice anywhere along the TGP coating. In this work, we report findings at ML = 34 mm and 49 mm. A Photomultiplier Tube (PMT) (R955HA, Hamamatsu) and a lens was used to detect the temperature-dependent luminescence decays. For both phosphors, the same 725±25 nm bandpass filter was mounted on the lens. The narrow band filter helps reduce the contribution of flame luminosity on the acquired phosphorescence decay signals.   In addition, a 295 nm long-pass filter was used to further block 266 nm excitation beam. A Tektronix MSO3054 oscilloscope was used to acquire the detected signals. Luminescence decays were processed on a single-shot basis to evaluate individual luminescence lifetimes. Calibration of the lifetimes with temperature was performed by monitoring the temperature of a heated TGP-coated aluminium bar (TGP-calibration target), equipped with a thermocouple and thermal insulation, which cooled to room temperature.

*2.3. Flame front (CH\*) imaging*

A high-speed camera (VEO 710L, Phantom), fitted with a 433±14 nm bandpass filter, was used to visualize the evolution of the flame front (CH*) within the crevice. The camera imaged through the front window (Fig. 1b) with a repetition rate of 2.5 kHz and exposure time of 400 μs. The imaging setup provided a ~ 176 μm/pixel resolution. A 400 nm long-pass filter was used to block scattered light from the 266 nm laser used for surface thermometry. Figure 1d shows an example CH* image describing the flame distribution in the crevice as the flame arrives at the ML. The green-dash lines in the figure show the vertical section where the flame position was tracked relative to the ML.

In Fig. 1d the CH* images show what appears to be an elongated or thick flame front according to the CH* signal. To understand the flame geometry in more detail, additional experiments were conducted to image the flame through the side window to get an orthogonal view of the flame within the crevice. Figure 2 shows an example of these flame front images at time *t* and at time *t*+Δ*t*  (800 μs later). The flame front can be described by a curved profile having a pointed tip in the middle of the narrow passage with its leading-edge at position *y*, while the flame closest to the wall (trailing-edge) sweeps over the wall location at *y*-Δ*y*. In our previous work [8,31], it was shown that the wall temperature abruptly increases when the flame closest to the wall sweeps over the TGP coating at the measurement location. Therefore, when imaging through the front window, it is expected that the sharp increase in wall temperature will not occur until the trailing edge of the flame interacts with the ML. This would result in a temporal lag between the time the flame's leading-edge reaches the ML and the time the wall temperature increases, since the latter is induced by the flame's trailing-edge. This is further discussed in Sect. 4.1.1 and Sect. 4.2.

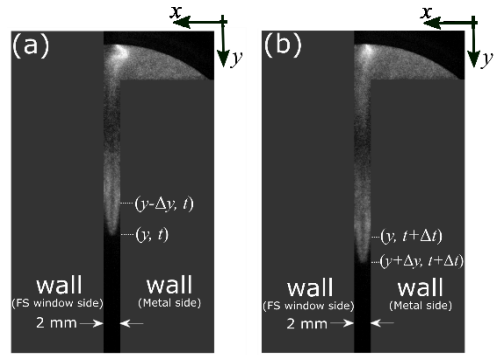

**Figure 2:** Selected images showing the flame front distribution in the crevice when imaged through the side windows of the FVC at (a) time, $t$, (b) time, $t+\Delta t$, ($\Delta t$ =800μs). The flame is imaged in the central plane of the crevice width (i.e. $z \approx 79$ mm) with a depth of field = 0.54 mm.

## 3.0. Thermographic Phosphors (TGPs)

This section evaluates the performance of ScVO for time-resolved $T_{wall}$ measurements in a FWI environment. ScVO was previously used for fluid thermometry at 10 Hz [33], which demonstrated impressive imaging capabilities in terms of precise (±0.4 °C (1σ)), single-shot temperature measurements. This work reports the first kHz rate measurements using ScVO for surface temperature measurements, and is applied here in a challenging FWI environment. ScVO is benchmarked against the performance of GGG, which has previously been employed for surface thermometry in FWI environments such as engines [30] and the FVC [8,31]. This section evaluates the luminescence properties for both phosphors and describes the performance of each phosphor in the two-wall FWI environment.

Figure 3 shows the room temperature emission spectra of GGG and ScVO. Both TGPs exhibit a broadband emission spectrum. The emission of GGG is attributed to the transition from the $^4T_2$ state to the $^3A_2$ ground state of $Cr^{3+}$ [39]; Ce, which is co-doped with Cr, does not influence the GGG's emission spectrum [40]. The emission of ScVO is credited to the $^3P_1 \rightarrow {}^1S_0$ and $^3P_0 \rightarrow {}^1S_0$ transitions of $Bi^{3+}$ [32]. The luminescence signals within the 725±25 nm wavelength range, which was specified by the spectral filter, is detected for temperature measurements.

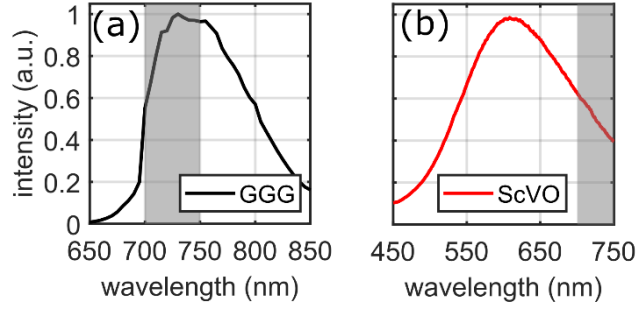

**Figure 3:** Room temperature emission spectrum of (a) GGG (data courtesy Phosphor Technology, UK), and (b) ScVO (data from [33] ). The grey region shows the spectral region of the bandpass filter used for temperature measurements.

### 3.1. TGP characterization

*3.1.1. Temperature-dependent luminescence lifetime*

Figure 4a shows the room temperature luminescence decay signal (average of 40 shots) for both ScVO and GGG. These decay signals were detected without the use of the 725±25 nm bandpass filter; however the scattered light from the 266 nm excitation beam was blocked from the PMT by using two 295 nm long-pass filters. As shown in Fig. 4a, the luminescence decay signal of ScVO is approximately 2 orders of magnitude shorter than GGG. Furthermore, the luminescence intensity of ScVO is higher. Quantitatively, the peak emission intensity of ScVO is 6 times higher than that of GGG. The higher luminescence intensity of ScVO is crucial for obtaining high signal-to-noise ratios (SNR).

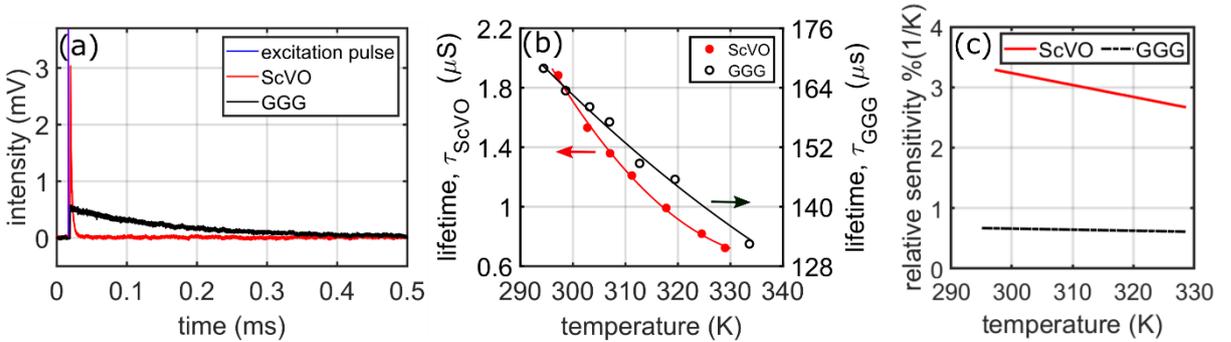

**Figure 4:** Luminescence properties; (a) room temperature luminescence decay signal, (b) temperature dependence of luminescence lifetime, (c) relative temperature sensitivity.

The room temperature luminescence lifetime ($\tau$) of a TGP provides an insight into the repetition rates at which measurements can be performed, in principle up to ~ 1⁄5$\tau$ Hz. After 5$\tau$ the luminescence signal is 0.7% of the peak signal. TGPs with $\tau < 200$ μs offer potential for kHz rates measurements. For the TGPs reported here, at 295 K, $\tau_{GGG}=$ 165 μs and at 297 K, $\tau_{ScVO} = 2$ μs. These decay times enable us to perform measurements at 1 kHz and 5 kHz, respectively. Fig. 4b shows the temperature dependence of $\tau$ obtained during a temperature calibration routine using a calibration target with the 725±25 nm bandpass filter in place. With increasing temperatures, $\tau$ for both TGPs becomes shorter, decreasing to 132 μs at 333 K for GGG and 0.73 μs at 329 K for ScVO. The calibration temperature range covered in Fig. 4b is relevant to changes in wall temperature measured for the various operating conditions reported in this paper (ΔT < 30 K). However, these selected TGPs have been shown to have temperature sensitivity up to 343 K [33] and 700 K [40] for ScVO and GGG, respectively.

Figure 4c shows the relative temperature sensitivity ( $(1/\tau)|(\delta\tau/\delta T)|$ ) derived from the calibration measurements. As shown, ScVO offers a sensitivity of 3.29 %/K at 297 K, which is nearly five times greater than GGG with 0.67%/K at 295 K. At higher temperatures of 328 K, the sensitivities of both phosphors decrease (2.67 %/K for ScVO and 0.61 %/K for GGG), but ScVO continues to exhibit a significantly higher sensitivity. At room temperature, the measurement precision (1σ) for ScVO is 0.2 K, which is a factor of three better than GGG at 0.6 K.

*3.1.2. Dependence of the measured temperature on excitation fluence*

For accurate temperature measurements from TGPs, changes in luminescence lifetimes should only occur due to changes in temperature. However, some TGPs are known to exhibit a dependence of luminescence properties (e.g. lifetime) on the excitation fluence [25]. Such dependence is attributed to laser-induced heating or effects related to the photo-physical properties of the TGP. To investigate how laser fluence effects would influence thermometry, tests were performed by probing each phosphor coating on the crevice wall within the FVC as illustrated in Fig. 1. The chamber was open to the ambient and did not operate with combustion during these tests. For these tests, both TGPs were excited at 1 kHz to maintain a similar laser fluence range. The dependence of the measured temperature on excitation fluence for the TGPs was then explored by measuring their respective τ under the conditions of varying excitation fluence. The laser fluence was varied using a waveplate and beam-splitter. The measured τ was converted to temperature using respective TGP calibration data obtained at a fluence of ~ 0.14 mJ/cm$^2$.

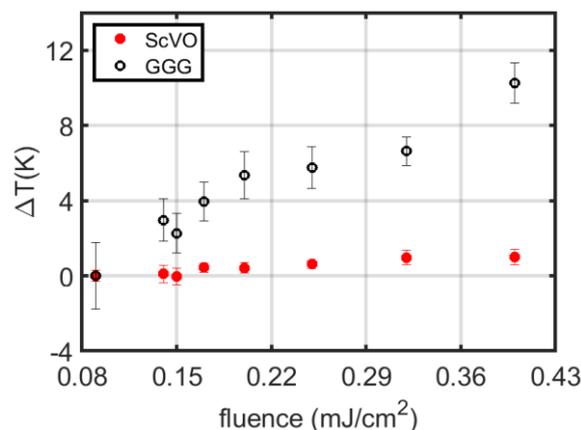

**Figure 5:** Cross-dependence of measured temperature on excitation fluence for ScVO and GGG. The error bars represent 1σ of the measured temperature. Varying excitation fluence was performed in random order.

Figure 5 shows the cross-dependence of measured temperature on excitation fluence. In Fig. 5, the measured temperature ($T$) is presented as a change in temperature ($\Delta T = T_{\text{(fluence)}} - T_{\text{(minimum fluence)}}$) for the selected TGPs. From the measurements, a strong dependence of excitation fluence on the measured temperature up to 11 K is observed for GGG for fluence in the range 0.08-0.43 mJ/cm$^2$. A similar fluence dependence was observed for GGG for the laser energy range of 0.04-0.2 mJ [30]. However, for ScVO only a minor temperature difference of less than 2 K is shown for the same 0.08-0.43 mJ/cm$^2$ fluence range. Abram et. al. [33] reports that variation in laser fluence across a range between 2 - 60 mJ/cm$^2$ for ScVO yields a temperature difference of ~ 3 K. This demonstrates that ScVO is advantageous over GGG in environments where excitation fluence can change due to experimental operating conditions.

### 3.2. Evaluation of TGPs for measurements in FVC

Similar to an IC engine, the operation of the FVC is characterized by variations in pressure and the presence of product gases from combustion, which can potentially attenuate light to various degrees. Stated briefly, the luminescence lifetimes (and measured surface temperature) of ScVO and GGG were insensitive to variations in gas pressure. This was confirmed by conducting experiments at static air pressures from 0.02 – 7.7 bar in the FVC. However, following the discussion in Sect. 3.1.2, it is important to investigate how changes in the excitation fluence influence thermometry in the FVC. To do this, surface temperature measurements were made on the metal wall in the crevice region of the FVC during combustion of a $\phi$ = 1.0 mixture, at $P_i$ =1 bar with CS = 2 mm. For these experiments, wall temperature ($T_{wall}$) measurements with ScVO was performed at 5 kHz, and with GGG at 1 kHz. The laser fluence at 5 kHz for ScVO was 0.05 mJ/cm$^2$, while the laser fluence at 1 kHz for GGG was 0.14 mJ/cm$^2$.

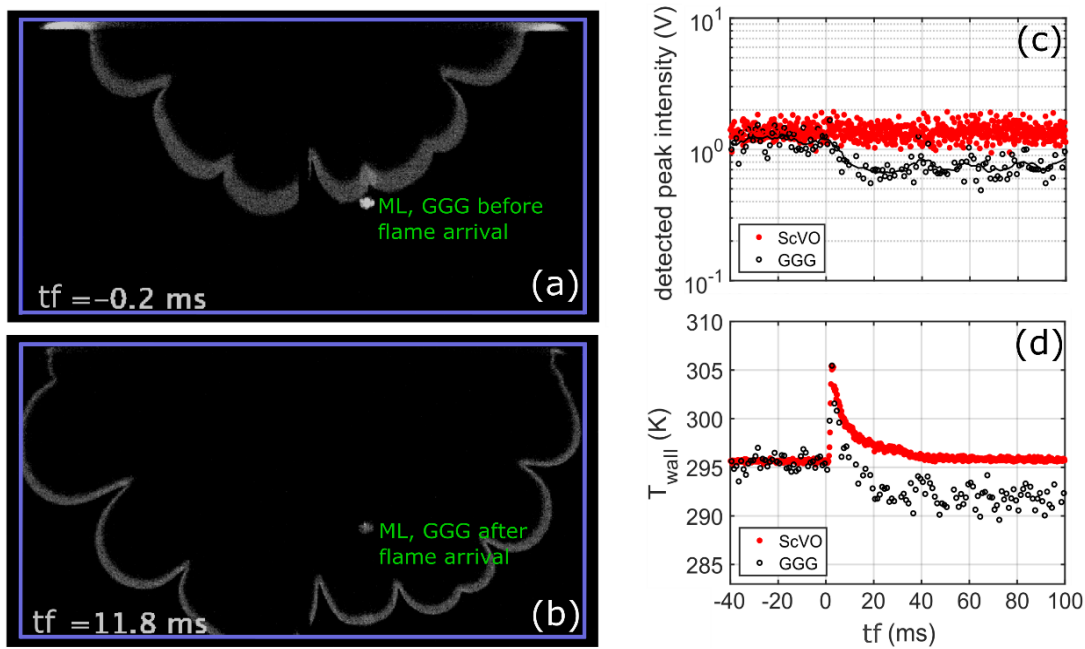

**Figure 6:** Image sequence of the flame in the crevice at reference times, (a) $tf$ = -0.2 ms and (b) $tf$ = 11.8 ms. The reference time $tf$ = 0 ms refers to the time the flame's leading edge reaches the ML. Time-history of (c) detected peak intensity, and (d) wall temperature ($T_{wall}$). Results reported for operation with $P_i$ = 1 bar, CS= 2 mm, and $\phi$ = 1.0.

Figures 6a and b show selected CH* images of the propagating flame in the crevice when GGG is probed for thermometry. The time, $t_f$, reported in these images is referenced to the time the flame's leading-edge arrives at the ML. The ML is seen on this image sequence due to the transmission of luminescence light through the 433±14 nm filter used for CH* imaging. For $t_f < 0$ ms (Fig 6a), the intensity of the GGG shown in the image sequence is noticeably brighter than after the flame has crossed the ML ($t_f > 0$ ms; Fig. 6b). This aspect is further shown in Fig. 6c, which shows the peak luminescence intensities detected for each TGP. It is shown that after the flame progresses beyond the ML ($t_f > 3$ ms), the peak luminescence decreases by ~ 32 % for GGG. For measurements conducted with ScVO, this decrease in peak luminescence is not observed.

Figure 6d presents the resulting $T_{wall}$ measurements derived from the detected luminescence decay signals from each TGP. Before the flame arrival at the ML ($t_f < 0$ ms), when the wall is exposed to unburned gases, there is no change in wall temperature. From time $0 < t_f < 3$ ms, the flame heats the wall and causes the abrupt rise in wall temperature. The maximum change in wall temperature ($\Delta T_{wall,max}$) recorded from ScVO and GGG is similar (~ 10 K). For $t_f > 3$ ms, the flame has passed the ML such that the wall is exposed to burned gas products. During this period, the wall temperature decreases exponentially with time. However, for GGG, $T_{wall}$ is observed to decrease to ~291 K, which is below its initial value of 295 K. This trend is not shown for ScVO, where $T_{wall}$ is shown to decrease and converge to its initial value with time.

It is clear that GGG struggles to accurately resolve $T_{wall}$ in post-flame regions, while this is not apparent for ScVO. While data is shown from a single experiment, all experiments are consistent with this trend. Due to the severe heat losses within the narrow crevice, it is expected that fuel will not oxidize completely. Thus, the product gases are likely to include unburned hydrocarbons, in addition to $CO_2$ and $H_2O$. At the conclusion of each experiment, although the gas contents of the chamber were partially evacuated through the dump valve, the effect of burned gas regions were evident in the chamber via visual inspection. Regions of burnt gas featured some opacity as glass components exposed to burned gases exhibited a faint layer of water condensation after the conclusion of each experiment. This is reasonable as gases experience rapid cooling near walls and during the expansion, such that the burned gases reach the dew point by the end of the exhaust event. All gases and water condensation were removed from the chamber via the vacuum pump, which sustained an absolute chamber pressure of 20 mbar for 1-2 minutes before the start of the next experiment. Phosphor measurements were highly repeatable for all tests. Thus, it is not expected that gas products or water condensation degraded phosphor performance throughout the experiment campaign.

It is believed that the product gases may be causing an optical attenuation that interferes with the GGG measurements. This attenuation can affect the laser excitation and/or the luminescence emission. It is plausible that absorption by, or scatter from, product gases could potentially attenuate the excitation fluence. As shown in Fig. 5, an attenuation of excitation fluence can lead to an appreciable temperature bias for GGG, while such a temperature bias would be nearly non-existent for ScVO. This explains the observed trends in Fig. 6d. Lower excitation fluences would also yield lower emission intensities. This effect is expected to occur for both TGPs as they were both excited by the same laser with a non-saturating excitation fluence, i.e., within a regime where the emission intensity varies linearly with excitation fluence. However, the data shown in Fig. 6c shows that the peak emission intensity decreases noticeably for GGG, but not for ScVO. This unique feature in emission intensity requires further investigation.

It is important to note that the PMT employed in this study operated within its linear regime. At the obtained luminescence levels, the peak PMT output current detected was < 0.05 mA for which the linearity is quoted to be < 2%. The decrease in emission intensity for GGG but not for ScVO remains an ongoing investigation. Nonetheless, it is clear that the GGG measurements exhibit a systematic error in the post-flame regime, while ScVO measurements do not exhibit this error. This systematic error associated with GGG can be reduced if the excitation fluence is significantly reduced [31]; however, this yields much lower SNR levels, hampering precise $T_{wall}$ measurements. Thus, this potential solution was not pursued in this work.

The evaluation between ScVO and GGG in Sect. 3 reveals that ScVO can offer notable advantages compared to GGG. Specifically, ScVO offers: (i) higher signal intensities, and thus higher SNR levels, (ii) higher temperature sensitivities, which provides improved temperature precision better than 0.2 K, (iii) opportunities for higher recording rates, demonstrated here at 5 kHz, and (iv) a negligible cross-dependence of the indicated temperature on the excitation laser fluence. The culmination of these advantages demonstrates that ScVO offers precise and reliable surface temperature measurements within the two-walled crevice during FWI. It is recognized that due to thermal quenching, ScVO has a limited temperature range (290 – 343 K, [33]) compared to GGG (290 – 700 K, [40]). However, for the application at hand, ScVO outperforms GGG in many aspects, particularly in post-flame environments.

**4.0. Results and discussion of measurements using ScVO**

In this section, ScVO is used to resolve wall temperature transients unique to flame and gas behaviour in narrow two-wall passages. Section 4.1.1 evaluates $T_{wall}$ and heat loss for variations in crevice spacing, while Sect. 4.1.2 resolves $T_{wall}$ transients in relation to flame dynamics. In Sect 4.2, wall heat flux is calculated to estimate flame quenching distances for flames burning or extinguishing at the ML.

**4.1. Demonstration**

*4.1.1. $T_{wall}$ transients accompanying variation in crevice spacing*

$T_{wall}$ measurements are performed in the crevice region for various crevice spacings (CS = 1.2, 2.0, and 3.5 mm) to investigate the rate of heat loss for various surface area to volume ratios. These experiments are performed with an initial chamber pressure of $P_i$ = 2 bar and $\phi$ = 1.0. Figure 7a shows individual pressure curves for each CS. As CS increases, the pressure curve broadens and exhibit higher peak pressures, which indicates higher heat release rates with increasing crevice spacing. These higher heat releases occur because a larger volume is available for gases to burn within the crevice region. CH* images show that a flame penetrates more easily into the crevice region for larger CS. As such, the flame consumes a larger portion of the crevice gas, yielding a higher heat release.

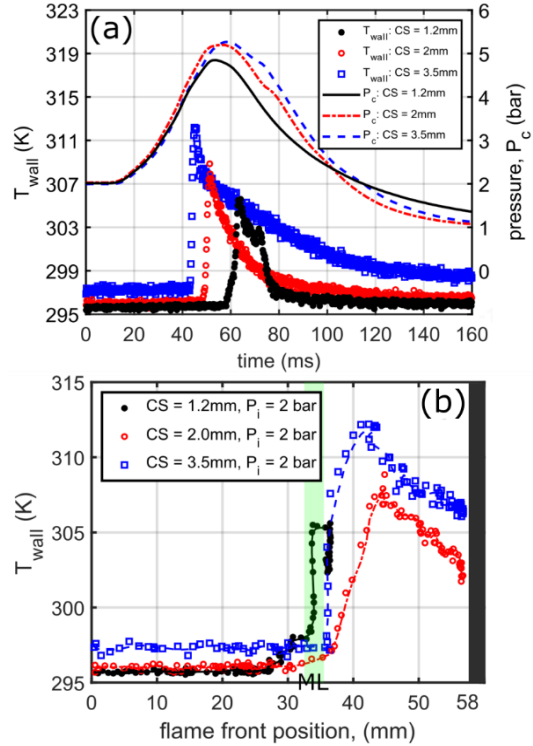

**Figure 7:** $T_{wall}$ measurements at $P_i$ =2 bar and $\phi$ = 1.0 under variable CS of 1.2, 2 and 3.5 mm. (a) Time-history of $T_{wall}$ and chamber pressure ($P_c$), (b) $T_{wall}$ as a function of flame front positions. ML = 34 mm ($\Delta y \sim$ 3 mm) is shown by the green patch.

Figure 7a reports the evolution of $T_{wall}$ for three individual experiments at each crevice spacing. These individual experiments, which correspond to the individual pressure curves in Fig. 7a, show qualitative trends consistent among each CS. $T_{wall}$ is measured at a distance of ML = 34 mm into the crevice. FWI causes an abrupt change in wall temperature, $\Delta T_{wall,max}$. The flame arrives earlier for larger CS, which is indicated by the timing of the temperature rise. Values of $\Delta T_{wall,max}$ in Fig. 7a correspond to 15K, 12K and 10K for CS of 3.5 mm, 2 mm, and 1.2 mm, respectively. For the cases of CS = 3.5 mm and 2 mm, $T_{wall}$ exhibits a pronounced exponential decay, indicating the rate at which the wall cools after the flame passes. For CS = 2 mm, the wall cools at a rate of 0.23 K/ms, which is approximately twice the value of 0.11 K/ms for CS = 3.5 mm. This demonstrates a greater rate of heat loss for the CS = 2mm case as the $SA/V$ is higher (1.02 mm$^{-1}$) than that for the CS = 3.5 mm case (0.59 mm$^{-1}$). For CS = 1.2 mm, $T_{wall}$ exhibits an initial decay, but then temporarily remains constant before exhibiting a rapid decay rate of 1.05 K/ms. This cooling rate is 4.6 and 9.5 times greater than the 2 mm and 3.5 mm spacings, while $SA/V$ changes by a factor of 1.65 and 2.86, respectively. This emphasizes the significance of heat loss with greater $SA/V$.

Figure 7b further describes the $T_{wall}$ behavior by reporting $T_{wall}$ as a function of flame front position. The 0 mm position is the position where the flame first enters the crevice. As discussed in Sect. 2, the flame position is defined as the leading edge of the flame identified in CH* images and is only evaluated in the location of the phosphor strip (green-dashed line in Fig. 1d). Since CH* was imaged at 2.5 kHz, the flame position is interpolated to match the 5 kHz $T_{wall}$ measurements. The data in Fig. 7b shows that the flames for CS = 2 mm and 3.5 mm penetrate well beyond ML = 34 mm and reach near the bottom of the crevice region. In these cases, $T_{wall}$ increases once the flame's leading-edge is several mm past the ML. As mentioned in Sect. 2, the flame's leading-edge corresponds to the flame tip (see Fig. 2), while the flame closest to the wall (trailing edge) is further upstream. Thus, $T_{wall}$ increases as the flame closest to the wall sweeps over the ML, which is typically 100's of μs after the flame tip passes the ML. This is shown as a "temporal lag" between flame location and $T_{wall,max}$ within Fig. 7b.

For the CS = 1.2 mm case, the flame propagates more slowly into the crevice region and does not penetrate the entire depth of the crevice. Instead, the flame penetrates to the ML and then remains relatively stationary. In this instance, $T_{wall}$ increases and reaches its maximum while the flame is within the ML, compared to the other cases where $T_{wall,max}$ is reached as the flame's leading-edge is beyond the ML. This occurs for the CS = 1.2 mm case because the distance between the flame tip and the flame closest to the wall (i.e. $\Delta y$ in Fig. 2) is smaller as the flame slows down in the crevice. This is evidenced from the CH* images, which exhibit a "narrowing" of the CH* signal as the flame slows down or stops in the crevice. Such CH* features, discussed further in Sects. 4.1.2 and 4.2, would indicate a less pronounced "temporal lag" between the time the flame arrives and the time $T_{wall}$ increases. Eventually, the flame penetrates just beyond the ML, at which point $T_{wall}$ decreases. Shortly afterwards, the flame recedes back into the ML where $T_{wall}$ increases (seen more clearly in Fig. 7a) and then the flame extinguishes in the vicinity of the ML. Further analysis of extinguishing flames at CS = 1.2 mm are presented in Sect. 4.2.

*4.1.2. $T_{wall}$ transients associated with flame dynamics*

CH* imaging has revealed that the flame in the crevice can be highly wrinkled. This is particularly true for higher pressures, where hydrodynamic instabilities can play a role in flame wrinkling [41,42]. The formation and evolution of flame wrinkles are highly transient and so is the $T_{wall}$ associated with the FWI of these features. This section describes these FWI transients and highlights the benefits of high precision $T_{wall}$ measurements at 5 kHz.

Figure 8 shows CH* and $T_{wall}$ measurements for a highly wrinkled flame propagating in the crevice. These experiments were recorded with $P_i$ = 2 bar and CS = 2 mm, where the flame penetrates into the entire crevice region. Figure 8a shows the overall view of the flame in the crevice at 44 ms, when the flame's leading-edge is near the ML. Figures 8b-e detail the evolution of a flame cusp interacting with the ML, where arrows are used to annotate the direction of flame movement observed in the high-speed images. Figure 8f reports the corresponding $T_{wall}$ at the ML during these events. The "narrowing" of the CH* signal, as described in Sect. 4.1.1, is shown within this image sequence. This narrowing is shown for flame cusp regions, where the flame is locally stationary compared to convex flame regions, which penetrate into the crevice and exhibit much thicker CH* regions.

As shown in Fig. 8, at 46 ms, the flame interacts with the ML, causing $T_{wall}$ to increase. $T_{wall}$ reaches a local maximum of 311.2 K at 48.5 ms just after the flame sweeps over the ML. $T_{wall}$ temporarily decreases before a region of the flame cusp interacts with the edge of the ML from 51-53 ms, at which point $T_{wall}$ is sustained at 308 K. The flame cusp temporarily moves away from the ML causing a drop in $T_{wall}$. At 56 ms, the flame cusp propagates upstream and penetrates through the ML, causing a second peak in $T_{wall}$ of 312.3 K at 57.5 ms. The flame on each side of the cusp eventually merge and the flame penetrates further into the crevice, at which time $T_{wall}$ exhibits an exponential decay as the wall cools.

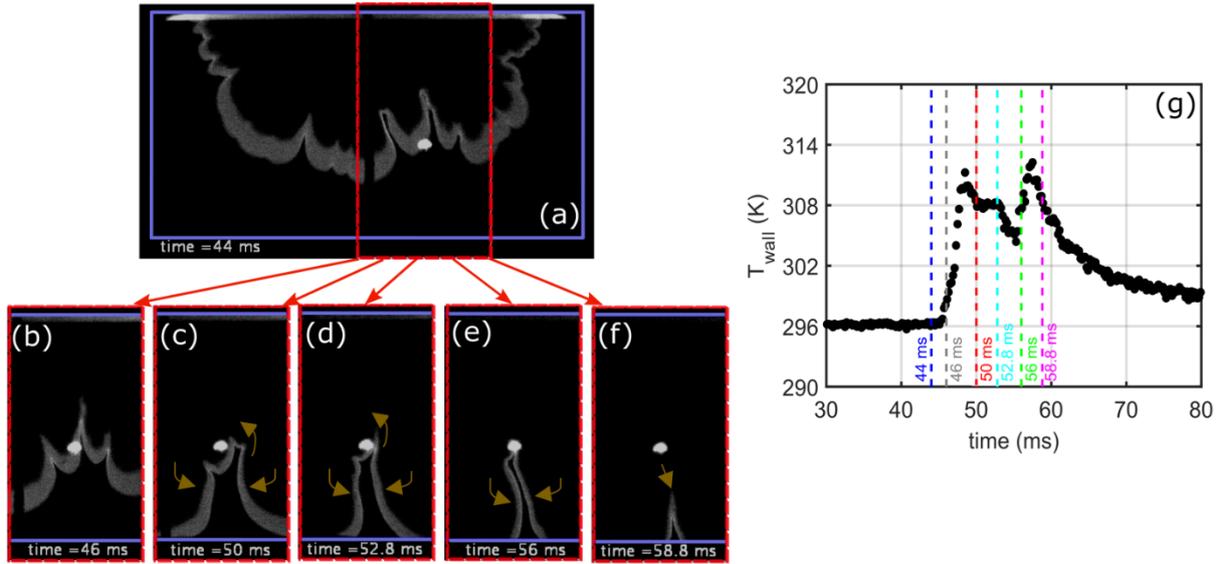

**Figure 8:** (a) – (f) Selected CH* images showing evolution of flame cusp with the ML. Operating conditions: $\phi$ = 1.1, $P_i$ =2 bar, and CS = 2mm. (g) Time-history of $T_{wall}$.

### 4.2. Evaluation of wall heat flux and flame quenching distance

This section capitalizes on the precise $T_{wall}$ measurements using the ScVO phosphor to derive measurements of wall heat flux ($Q_w$) and flame quenching distance ($\delta_q$) within the two-walled crevice. These quantities are evaluated for critical FWI conditions in which a flame extinguishes within the crevice. The crevice spacing of CS = 1.2 mm is chosen for this operation as it yields high surface area to volume ratios, for which heat loss is considerable and leads to partial or complete flame quenching. Experiments are performed for two initial operating pressures of $P_i$ = 1.4 bar and 2.0 bar. The flame propagates further into the crevice under higher pressures. The phosphor location is therefore placed at ML = 34 mm and 49 mm into the crevice for the 1.4 bar and 2.0 bar initial pressures, respectively. These are the average distances the flame reaches before extinguishing. This section first describes the flame penetration and corresponding $T_{wall}$ for these conditions, after which $Q_w$ and $\delta_q$ are derived and presented. For $T_{wall}$ measurements presented in this section, a 607±18 nm bandpass filter and a separate calibration was used. This filter is centred near the peak emission intensity of ScVO, which further improved luminescence signal levels detected by the PMT.

In this section, it is important to emphasize that the quantities derived from $T_{wall}$ measurements are unique to the FWI event that occurs at the ML. As described in Sect. 4.1.2, the flame-wall dynamics vary spatially and temporally. Moreover, the specific FWI event at the ML varies for each experiment. The aim of this section is to report *local* flame-wall quantities that depict the FWI at the ML. We also report qualitative trends in quenching distance ($\delta_q$) for flames that actively burn near the ML and for flames that extinguish near the ML. In order to estimate $\delta_q$, the maximum wall heat flux ($Q_{w,max}$) is required. We calculate $Q_{w,max}$ at the ML for the aforementioned flame variations. For extinguishing flames, it is understood that $Q_{w,max}$ will be lower than flames that actively burn at the ML. However, a measure of $Q_{w,max}$ at the ML provides useful information of the flame behaviour at that location. Thus, it is important to understand that the values reported are not universal values, but rather values *local* to the specific FWI event occurring at the ML.

It is also important to distinguish between the terminology "quench" and "extinguish". In this work, we estimate the quenching distance, which is defined as the flame distance from the wall. This occurs because the flame loses heat to the wall (strain can also play a role [43,44]), such that chemical reactions are quenched and a flame can only burn a certain distance from the wall. When quenching distance becomes extreme, the flame will extinguish altogether for which heat release eventually ceases. We refer to this scenario as "extinguish".

### 4.2.1. Flame penetration depth and $\Delta T_{wall,max}$

Figure 9 describes the flame penetration into the crevice and the flame's interaction with the ML for select cases at each initial pressure. Figure 9a shows the time history of the flame's leading-edge within the crevice for the selected cases. As previously mentioned, the flame's location is evaluated in-line with the phosphor coating (green-dashed line, Fig. 1d). Figures 9b-c show still images of the flame for each case as the flame reaches its maximum distance into the crevice. As mentioned previously, the CH* images show a thin layer representing the flame front, compared to the thicker layers that exist as the flame is actively propagating into the crevice (e.g. Fig. 6a and Fig. 8a). This suggests that the flame tip and flame closest to the wall, as described in Fig. 2, are at similar locations as the flame slows down in the crevice.

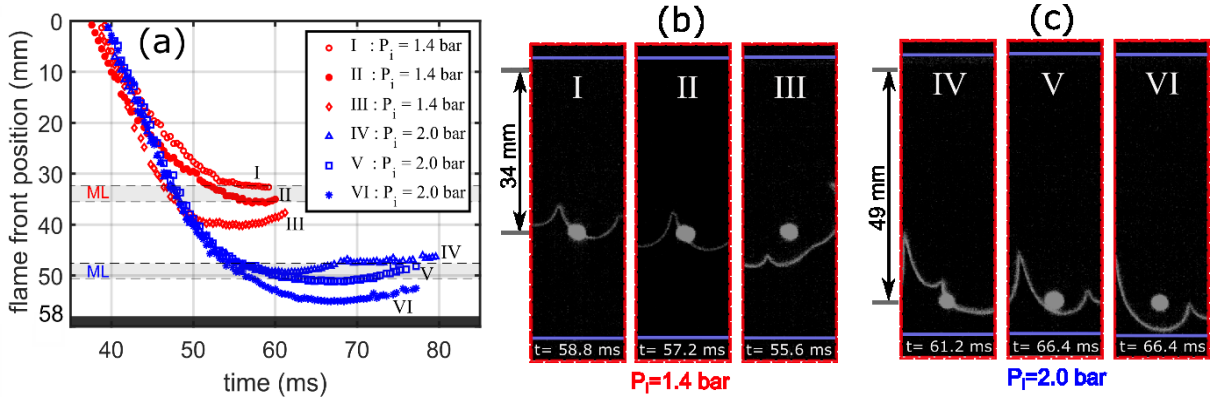

**Figure 9:** (a) Time-history of flame position for selected cases at $P_i$ = 1.4 and 2.0 bar with CS = 1.2 mm. ML is shown as grey outline. CH* images of the flame front relative to the ML, when the flame is at the furthest position, are shown in (b) $P_i$ = 1.4 bar, ML = 34 mm, and (c) $P_i$ = 2.0 bar, ML = 49 mm.

As shown in Fig. 9a, the flame propagates into the crevice, and for the first 20 mm the flame position exhibits a linear relationship with time. This behaviour is consistent for both $P_i$ conditions. An approximate propagation speed can be derived from the position-time relationship and, on average, is 3.1 m/s and 4.0 m/s for $P_i$ = 1.4 bar and 2.0 bar cases, respectively. There is a variation in flame penetration depth amongst the cases shown. For cases I and II ($P_i$ = 1.4 bar), the flame slows down after 20 mm into the crevice and the flame stops while it is in contact with the ML. For case III, the flame penetrates further into the crevice and stops beyond the ML. For the $P_i$ = 2.0 bar cases, the flame penetrates much further into the crevice, with cases IV and V stopping within the ML at 49 mm, while case VI penetrates further downstream. For cases III to VI, the flame position moves upstream after it reaches its maximum distance. This occurs as the gases in the chamber undergo expansion/exhaust as the dump-valve is activated and the chamber pressure decreases. The last data points in Fig. 9a indicate the flame location and time at which the flame extinguishes.

Figure 10 reports the maximum wall temperature change ($\Delta T_{wall,max}$) as a function of maximum flame penetration depth. As shown previously, $\Delta T_{wall,max}$ occurs as the flame interacts with the ML. Data is shown for all experiments performed at these conditions, and cases I – VI are identified in Fig. 10. Experiments for which the flame did not arrive at the ML are reported with $\Delta T_{wall,max} = 0$ K. As shown in Fig. 10, $\Delta T_{wall,max}$ is lowest for flames that extinguish within the ML. There is some variation in $\Delta T_{wall,max}$ for these flames, and this is inherent to the specific FWI at the ML for each case. However, all cases where a flame extinguishes at the ML exhibit $\Delta T_{wall,max}$ less than 7 K for both $P_i$ conditions.

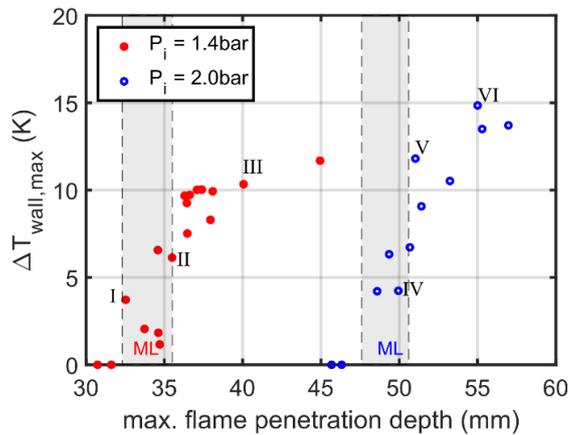

**Figure 10:** Variation of the maximum change in wall temperature ($\Delta T_{wall,max}$) versus maximum flame penetration depth. Grey outline represents the ML.

For cases where the flame penetrates beyond the ML, $\Delta T_{wall,max}$ is greater and is shown to increase with increasing penetration depth. Again, there is some variation in this trend, which is due to the specific nature of the FWI event at the ML. For example, case V shows a relatively high $\Delta T_{wall,max}$ compared to other cases which penetrate a similar distance. As shown in Fig. 9a, for case V the flame is in contact with the ML for nearly 15 ms, which is much longer than other cases and likely contributes to the higher $\Delta T_{wall,max}$. As shown in Fig. 9a, flames penetrating well beyond the ML typically pass through the ML more quickly. This would suggest that the flame is actively burning past the ML as opposed to slower or stationary flames that show signs of weakening via declining CH* intensities. These "stronger" flames are anticipated to have higher flame temperatures and burn closer to the wall than "weaker" flames which slow down and extinguish near the ML. Although it is not possible to measure flame temperature with these measurements, the estimated flame distance from the wall is evaluated as quenching distances in the next section.

*4.2.2. Wall heat flux and estimated quenching distance*

Boust et al. [36] demonstrated an effective formulation to derive quenching distance from relatively simple measurements of wall heat flux. Their formulation was derived from first principles by applying an energy balance to a flame. They applied this derivation to calculate the quenching distance ($\delta_q$) for a *single-wall* configuration for both head-on and side-wall quenching. In this section, we cautiously apply this derivation to *estimate* $\delta_q$ for our *two-wall* configuration with side-wall quenching. We recognize that differences in flame quenching exist between single- and two-wall configurations. Most notably, the quenching distance and relevant Peclet number ($Pe = \delta_q/\delta_F$; $\delta_F$ = flame thickness; $\delta_F \equiv$ thickness of preheat zone ($\delta_p$)) are typically larger for two-walls than for single-walls [45–47]. However, the derivation from first principles is very similar for both cases and is described below.

*4.2.2.1 Derivation*

A schematic of the flame in the two-walled crevice is shown in Fig. 11. An energy balance applied to the flame front states that the flame power (i.e. available chemical energy) is distributed between the thermal power of the reaction zone and the heat lost through the preheat zone and to the surroundings:

$$\underbrace{\rho_u S_u Y_{fuel} \Delta H}_{flame\ power\ =\ Q_\Sigma} = \underbrace{\rho_u S_u c_p (T_F - T_u)}_{thermal\ power\ rxn\ zone} + \underbrace{Q_{loss}}_{heat\ loss} \quad (1)$$

In Equation 1, $\rho_u$ is the unburned gas density, $S_u$ is the unstretched laminar flame speed, $Y_{fuel}$ if the fuel mass fraction of the unburned mixture, $\Delta H$ is the enthalpy of combustion, $c_p$ is the specific heat capacity, $T_F$ is the flame temperature, and $T_u$ is the unburned gas temperature. Equation 1 is argued to hold true as long as a flame is actively burning, such that the flame power ($Q_\Sigma$) exists. This is true for both single- and two-wall configurations, in addition to situations without a wall.

The thickness of the preheat zone, $\delta_p$, is modelled as conduction through the unburned gases:

$$\delta_p = \frac{\lambda}{\rho_u S_u c_p} \quad (2)$$

Where $\lambda$ is the thermal conductivity of the unburned gas. At the wall, $Q_{loss}$ is equal to the heat lost at the wall, which can be directly measured as the wall heat flux ($Q_w$). A thin layer of gas, denoted as the quenching distance ($\delta_q$), separates the flame from the wall. The wall heat flux through $\delta_q$ can be expressed as:

$$Q_w = \lambda \left(\frac{\partial T}{\partial x}\right)\bigg|_{0<x<\delta_q} = \lambda \frac{(T_F - T_w)}{\delta_q} \quad (3)$$

A simplifying assumption is made to approximate $T_F - T_w \approx T_F - T_u$ at the wall. Although the unburned gas temperature close to the wall can deviate from the wall temperature during FWI [8], this deviation is small compared to the magnitude of $T_F - T_u$. Thus, this assumption is considered reasonable and is equally applicable for single- and two-wall configurations.

The challenge with Equations 1 and 3 is that it is difficult to determine $T_F$ during transient flame quenching. This is true for both single- and two-wall configurations. Thus, it is desired to remove $T_F$ by substitution. Solving for $T_F - T_w$ in Equation 3 and substituting this expression into Equation 1 gives:

$$\rho_u S_u Y_{fuel} \Delta H = \frac{\rho_u S_u c_p}{\lambda} \delta_q Q_w + Q_w \qquad (4)$$

Equation 4 can be further simplified to:

$$\frac{\rho_u S_u Y_{fuel} \Delta H}{Q_w} = \frac{\delta_q}{\delta_p} + 1 \qquad (5)$$

$$\delta_q = \frac{\rho_u S_u Y_{fuel} \Delta H}{Q_w} \delta_P - \delta_P \qquad (6)$$

$$\delta_q = \frac{Y_{fuel} \Delta H \lambda}{Q_w c_p} - \frac{\lambda}{\rho_u S_u c_p} \qquad (7)$$

Equation 7 provides a rather simple expression in which $\delta_q$ can be determined from the mixture properties and direct measurements of $Q_w$. Boust et al. [36] demonstrated good agreement (within 20%) for $\delta_q$ values determined from Equation 7 and those determined from flame imaging.

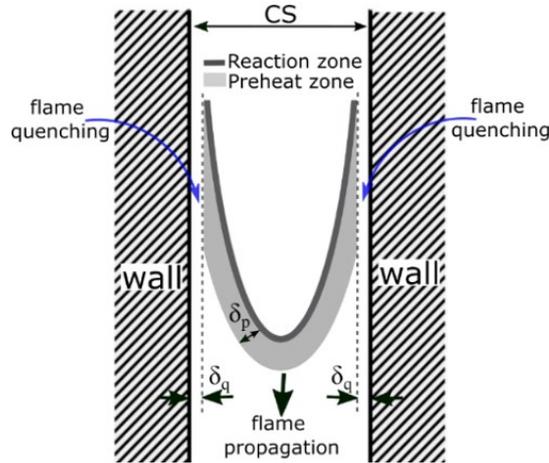

**Figure 11:** Schematic of sidewall quenching in a narrow two-wall passage as indicated in Fig. 2.

The *two-wall* side-wall quenching configuration in this work exhibits differences to the *single-wall* side-wall quenching configuration in [36]. For example, in the narrow passage, because the flame sweeps across two walls, there would be two $\delta_q$ values associated with each wall (see Fig. 11). $\delta_q$ on each wall is anticipated to be similar if the wall materials are the same. For the crevice region in the chamber, one wall is comprised of fused-silica and the other is stainless-steel. In this work, we measure $Q_w$ on the stainless-steel wall, thus only having the capability to measure $\delta_q$ for one of the two walls. Fused-silica has a lower thermal conductivity than stainless steel, such that $\delta_q$ is anticipated to be larger for the stainless-steel wall.

It is also important to emphasize that the true flame displacement speed, $S_d$, will deviate from the unstretched laminar flame speed, $S_u$. The above analysis utilizes $S_u$ since this is a property that can be easily obtained from the literature. Indeed, researchers have shown that laminar $S_d$ can deviate from $S_u$ by up to 25% (e.g.,[48,49]). Such deviations can occur due to flame stretch or from velocity differences across the flame front. Those findings have been reported for spherical flame geometries; they have not been reported for flames in narrow channels. Measurements elucidating the differences between laminar $S_d$ and $S_u$ within complex wall-bounded flows are rare. It would require measurements of flow velocity and a representative flame contour (e.g., [50]). Such measurement would be incredibly challenging within a two-walled crevice. Boust et al. [36] argued that flame stretch has a secondary effect on $S_u$ compared to thermal losses for single-wall side-wall quenching. Since it is not possible to obtain an accurate $S_d$ value, which can be substituted for $S_u$ in Equation 7, a sensitivity analysis is performed to evaluate the effect of $S_u$ on $\delta_q$. $\delta_q$ is recalculated using a ± 40% deviation from the nominal $S_u$ value. In Sect. 4.2.2.2, it is shown that this 40% deviation in $S_u$ only amounts to a 1-8% deviation in $\delta_q$. This is because the first term on the right side of Equation 7, which is independent of $S_u$, contributes over 90% of the total value of $\delta_q$. As such, deviations in $S_u$ will have a minor effect on the calculated $\delta_q$ values.

Although Equation. 7 represents a simplified derivation, we argue that this expression is suitable to provide a first-order approximation of $\delta_q$. As such, Equation 7 is adopted to calculate $\delta_q$ for the stainless-steel wall in the crevice. However, due to the challenges explained above, the reported values of $\delta_q$ in this work are referred to as *estimates* at the ML. This is true when the flame is actively burning near the ML. There are additional limitations for extinguishing flames, which are discussed in Sect. 4.2.2.2. We therefore restrict the discussions of $\delta_q$ to qualitative trends only.

4.2.2.2 Evaluation of $Q_w$ and $\delta_q$

The value of $Q_w$ in Equation 7, is determined from $T_{wall}$ measurements from ScVO. This formulation is based on the time-dependent surface temperature of a thermally semi-infinite solid. This is treated as a one-dimensional transient heat conduction problem, which uses a Duhamel integral that provides the time evolution of $Q_w$ given by [51,52].

$$Q_w(t) = \sqrt{\frac{\rho C_{p,w} k}{\pi}} \int_0^t \frac{[dT_{wall}(\bar{\tau})/d\bar{\tau}]}{\sqrt{t-\bar{\tau}}} d\bar{\tau} \qquad (8)$$

Here $\rho$, $C_{p,w}$ and $k$ are the density, specific heat capacity, and thermal conductivity of the wall material (stainless-steel), $t$ is time, and $\bar{\tau}$ is the time variable of integration.

Figure 12a shows the temporal evolution of the estimated $Q_w$ (based on Equation 8) and the measured $T_{wall}$ for a selected case at $P_i$ = 2.0 bar and CS = 1.2 mm. $T_{wall}$ measurements can exhibit a small amount of noise as shown in Fig. 12. A 5-point moving average (1ms interval) is applied to $T_{wall}$ to reduce this noise in the calculation of $Q_w(t)$. In Fig. 12, similar to $T_{wall}$, $Q_w$ exhibits an abrupt increase as the flame interacts with the wall, after which $Q_w$ shows a rapid decrease. The temporal evolution of $T_{wall}$ and $Q_w$ reported here are similar to those obtained in a fired IC engine [53]. The maximum wall heat flux corresponds to the heat flux at quenching, when the flame is closest to the wall [36,54]. Therefore, $Q_{w,max}$ is used as the value of $Q_w$ in Equation 7.

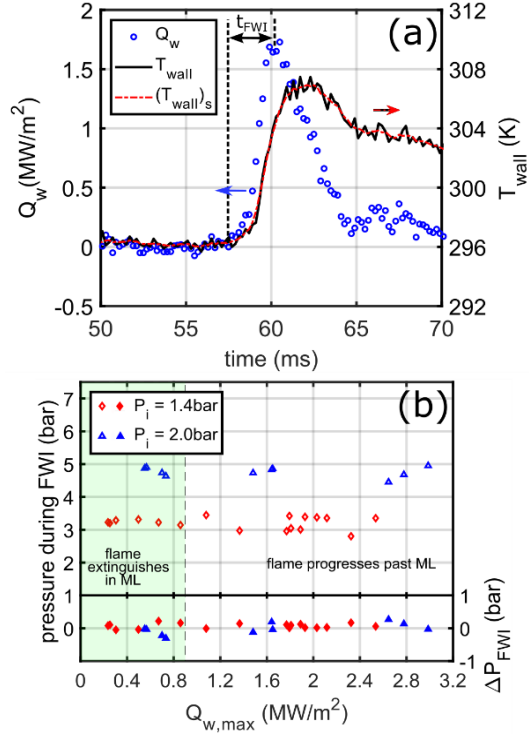

**Figure 12:** (a) Time-history of $Q_w$ and $T_{wall}$. $(T_{wall})_s$ is the 5-point moving mean (1 ms interval) of $T_{wall}$ used to calculate $Q_w$. (b) Pressure versus wall heat flux at quenching for flame quenching during FWI.

In applying Equation 7 for our analysis, all variables are considered at the thermodynamic conditions when the flame is interacting with the ML. Although the pressure is transient, the time in which the flame interacts with the ML ($t_{FWI}$) ranges from 1-6 ms. $t_{FWI}$ is identified within Fig. 12a and is the time during which $Q_w$ changes abruptly from 0 to $Q_{w,max}$. Figure 12b shows the pressure and change in pressure during FWI at the ML. The pressure during this short time duration changes, on average, by 0.05 bar.

Figure 13a shows the findings of $\delta_q$ in relation to $Q_{w,max}$ as well as maximum flame depth in the crevice. Data is shown for all experiments performed at $P_i$ = 1.4 bar and 2.0 bar with CS = 1.2 mm. Cases I to VI are identified within Fig. 13. The simultaneous flame front imaging and $T_{wall}$ measurements allow us to observe a correlation between $\delta_q$ and the flame penetration relative to the ML. In this analysis, we distinguish between flames that penetrate past the ML and flames that extinguish near the ML. Data points for the latter are shown in the green regions highlighted in Fig. 13. For flames that penetrate well beyond the ML, we believe that the flame actively burns while interacting with the ML. This is evidenced by the higher $\Delta T_{wall}$ shown in Fig. 10, where $\Delta T_{wall}$ is similar in value when the ML is further upstream the crevice (~12 K for $P_i$ = 1.4 bar; ~15 K for $P_i$ = 2.0 bar). As such, the expression for flame power ($Q_\Sigma$) is expected to be sufficient, for which Equation 1 can be applied. For cases where the flame weakens and extinguishes at the ML, the expression for $Q_\Sigma$ is not expected to hold; $Q_\Sigma$ is expected to be less for extinguishing flames. Thus, Equation 7 is anticipated to produce larger inaccuracies in $\delta_q$ for flames extinguishing near the ML.

Given these limitations, we focus on quantitative values of $Q_{w,max}$, but the *qualitative* variation in $\delta_q$. We also test the hypothesis that if $\delta_q$ is equal on both walls, then we would expect a flame to pass through the ML if $\delta_q$ < 0.5×CS, but we would expect the flame to extinguish near the ML if $\delta_q$ > 0.5×CS. We recognize the different wall materials will yield different $\delta_q$ values for each wall, but we expect this to be less significant for the means of testing this hypothesis.

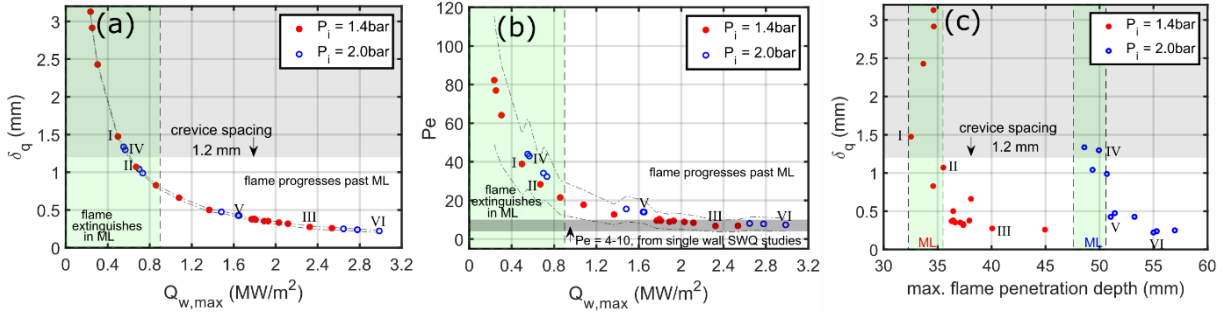

**Figure 13:** (a) Variation of quenching distance ($\delta_q$) versus wall heat flux at quenching ($Q_{w,max}$), (b) Variation of Peclet number versus $Q_{w,max}$, (c) Variation of $\delta_q$ versus maximum flame penetration depth. Dotted lines in (a) and (b) represent the upper and lower bounds of $\delta_q$ and Pe due to a ± 40% variation in $S_u$. This sensitivity analysis shows that $\delta_q$ estimated from Equation 7 is not very sensitive to $S_u$, because the first term in Equation 7, which is independent of $S_u$, dominates.

The data in Fig. 13a shows that flames penetrating beyond the ML yield $Q_w$ values between 1.0 – 3.0 MW/m², which correspond to estimated $\delta_q$ values between 0.66-0.22 mm. These values are all smaller than half of the crevice spacing, which supports the above hypothesis. Figure 13b shows the Peclet number (Pe) calculated from the estimated $\delta_q$. For flames penetrating beyond the ML, Pe ranges from 6 – 18. These values are similar to those reported for single-wall SWQ flames, Pe = 4-10 [36,55,56]. Figure 13a also shows that cases III and VI are among the lowest $\delta_q$ values estimated. Figure 13c reveals that these cases exhibit a flame that extends beyond the majority of cases within each dataset. This observation supports the argument that a flame with a smaller $\delta_q$ is able to penetrate further within the crevice before extinguishing.

For flames that extinguish near the ML, Fig. 13a shows that $Q_w$ remains below 0.9 MW/m² corresponding to $\delta_q$ values greater than 0.7 mm. That is, $\delta_q$ > 0.5×CS, which agrees with the said hypothesis. At $Q_w$ values lower than 0.55 MW/m², $\delta_q$ values are greater than CS. Such values of $\delta_q$ are not physically possible, emphasizing that the Equation 7 is not suitable to provide a realistic value for $\delta_q$ as the flame begins to extinguish. Again, this is likely due to the insufficient expression for $Q_\Sigma$ in Equation 1. Measurements of flame temperature while the flame is extinguishing would better enable an estimate of $\delta_q$ within Equation 1. This, however, is beyond the scope of this work.

It is also shown that Pe = 20 – 80 for the extinguishing flames. Values greater than Pe = 40 correspond to unrealistic $\delta_q$ values (i.e., $\delta_q$ > CS). The Pe values associated with $\delta_q$ < CS are consistent with those reported in the literature for extinguishing flames within two-wall configurations (Pe = 15 – 50 [45–47]). The consistent trends are encouraging despite the limitations in the $\delta_q$ estimation.

The ± 40% variation in $S_u$, as described in Sect. 4.2.2.1, is shown by the dotted lines in Fig. 13 a and b. These lines represent the upper and lower bounds of $\delta_q$ and Pe due to the ± 40% $S_u$ variation. As described earlier, $S_u$ has a minor effect on $\delta_q$ with values only deviating by 1-8%. $S_u$ has a larger effect on Peclet number because Pe is more approximately inversely proportional to $S_u$. With this small influence of $S_u$ on $\delta_q$, we argue that Equation 7 can sufficiently provide a qualitative $\delta_q$ despite not having a 100% accurate value for $S_u$. Moreover, although values of $\delta_q$ are considered qualitative, the $T_{wall}$ measurements applied to Equation 7 is able to distinguish between (i) flames with $\delta_q$ small enough to allow the flame to burn past the ML, and (ii) flames that exhibit severe quenching at the wall, which eventually leads to flame extinguishment at the ML.

## 5.0. Conclusions

Wall temperature measurements in a designated narrow two-wall passage (crevice) of an FVC have been performed at 5 kHz using phosphor thermometry. These measurements were combined simultaneously with flame front (CH*) imaging to capture how the spatiotemporal features of the flame influence the local surface temperature in the crevice. Two thermographic phosphors (TGPs), $Gd_3Ga_5O_{12}$:Cr,Ce; (GGG) and $ScVO_4$:$Bi^{3+}$ (ScVO), suitable for high-speed surface thermometry were characterized. ScVO was shown to be well suited for quantitative and high-fidelity measurements in the FVC due to its higher SNR, high-temperature sensitivity and precision, and its negligible susceptibility to attenuation of the excitation laser fluence by combustion product gases.

A demonstration of the time-resolved and precise crevice surface thermometry using ScVO was conducted. The rate of heat loss for various surface area to volume ratios was investigated for various crevice spacing (CS) with the results indicating that heat is lost more rapidly for narrower CS. Furthermore, the high precision wall temperature ($T_{wall}$) measurements at 5 kHz allowed to uniquely capture $T_{wall}$ transients associated with the highly transient formation and evolution of flame wrinkles within the two-walled crevice. Finally, by taking advantage of the precise $T_{wall}$ measurements with a crevice spacing of 1.2 mm, measurements of wall heat flux ($Q_w$) and quenching distance ($\delta_q$) were derived for FWI conditions in which a flame extinguishes at the measurement location (ML) or actively burns past the ML. Flames exhibiting higher $Q_w$ with $\delta_q < 0.5 \times CS$, actively burn past the ML and penetrate further into the crevice before extinguishing. Flames with lower $Q_w$ exhibit severe quenching at the wall, leading to flame extinguishment at the ML.

Investigating FWI for surfaces at higher temperature is the focus of future work. To do this in a time-resolved manner, a TGP that is sensitive at temperatures above 340 K will be selected. Future work will also extend high-speed phosphor measurements to 2D imaging, which will resolve flame-wall dynamics with an appropriate spatial resolution. Exploitation of various excitation schemes (including multi-photon excitation) as described in [57] may also be explored in flame environments.

## Acknowledgements

We gratefully acknowledge funding from the European Research Council (ERC grant # 759546) and EPSRC (EP/P020593/1, EP/P001661/1).